\documentclass[floatfix,aps,twocolumn,showpacs,preprintnumbers,amsmath,amssymb]{revtex4-1}

\usepackage[applemac]{inputenc}
\usepackage[T1]{fontenc}
\usepackage{amsfonts, amssymb, amsmath, dsfont}
\usepackage{graphicx}
\usepackage{lipsum}
\usepackage{float}
\usepackage{color}

\newcommand{\boF}{\mathbf{F}}
\newcommand{\PI}{\mathbf{\Pi}}
\newcommand{\boPhi}{\mathbf{\Phi}}

\newcommand{\bE}{\mathbf{E}}
\newcommand{\bcalE}{\boldsymbol{\mathcal{E}}}

\newcommand{\bH}{\mathbf{H}}
\newcommand{\bcalH}{\boldsymbol{\mathcal{H}}}

\newcommand{\bor}{\mathbf{r}}

\newcommand{\tk}{\tilde{k}}
\newcommand{\tq}{\tilde{q}}

\def\eps{\varepsilon}
\def\Re{\mathrm{Re}}
\def\Im{\mathrm{Im}}

\begin{document}

\title{Chiral near fields generated from plasmonic optical lattices}

\author{Antoine Canaguier-Durand}
\affiliation{ISIS \& icFRC, University of Strasbourg and CNRS, 8 all\'{e}e Gaspard Monge, 67000 Strasbourg, France.}
\author{Cyriaque Genet}
\altaffiliation{Corresponding author: genet@unistra.fr}
\affiliation{ISIS \& icFRC, University of Strasbourg and CNRS, 8 all\'{e}e Gaspard Monge, 67000 Strasbourg, France.}

\begin{abstract}
Plasmonic fields are usually considered non-chiral because of the transverse magnetic polarization of surface plasmon modes. We however show here that an optical lattice created from the intersection of two coherent surface plasmons propagating on a smooth metal film can generate optical chirality in the interfering near field. We reveal in particular the emergence of plasmonic potentials relevant to the generation of near-field chiral forces. This draws promising perspectives for performing enantiomeric separation schemes within the near field. 
\end{abstract}

\maketitle

\section*{Introduction}
 The capacity to extract clear stereochemical informations on chiral molecules is essential to a large number of analytical chemistry methods \cite{BarronBook}.
Recently, it has been emphasized that the rate of absorption in molecular chirality is controlled by the density of chirality of the irradiating light field, leading to the idea that ``superchiral'' fields can enhance chiroptical sensitivity levels \cite{TangScience2011,HendryNatNano2010,RheeChemSci2013}. \\
\indent These issues have revitalized the interest in optical chirality \cite{lipkin1964existence,tang2010optical,bliokh2011characterizing,coles2012chirality,cameron2013optical}. In particular, optical chiral forces described recently not only involve the density of chirality of the light field, as for chiral spectroscopy, but also its flow of chirality \cite{guzatov2011chiral,shang2013analysis,tkachenko2013spin,ding2013realizing,reid2013efficient,canaguier2013mechanical,cameron2014discriminatory,cameron2014diffraction,tkachenko2014optofluidic,wang2014lateral}. When chirality of light is coupled to chirality of matter, new possibilities for optical enantioseparation have been described, emphasizing that both concepts of chirality density and flow must be included in a comprehensive analysis of chiroptical schemes \cite{canaguier2013mechanical,cameron2014discriminatory,cameron2014diffraction,tkachenko2014optofluidic}.  \\
\indent In this context, surface plasmons (SPs) excited on a metal film are interesting: through their strong amplitude and phase gradients, they give rise to very efficient momentum transfers enabling direct object manipulations together with pulling effects that maintain particles in the near field \cite{garces2006extended,wang2009propulsion,juan2011plasmon,cuche2012brownian,cuche2013sorting,chen2013transport}. But while SPs are characterized by chirality flows due to their intrinsic spinning nature, plasmonic chiral densities identically vanish because of the transverse magnetic polarization of SPs \cite{bliokh2012transverse}. From a chirality perspective, SPs turn out therefore disappointing, explaining why spectroscopists have focused their attention on {\it localized} surface plasmon modes (LSP) that lead to strong chiral densities through high local field enhancements \cite{abdulrahman2012induced,DavisPhysRevB2013,SchaferlingPRX2012,lin2014slant}. Exploiting such localized excitations have allowed improving sensitivity levels in the context of molecular chiral spectroscopy \cite{TangScience2011,HendryNatNano2010,RheeChemSci2013}.   \\
\indent In this article, we take a different view by realizing that a plasmonic optical lattice, formed by the intersection of two coherent surface plasmons propagating perpendicularly with each other on a smooth metal film, provides all necessary ingredients for performing chiroptical schemes in the near field. As we will show, such a near-field optical lattice indeed gives rise to an inhomogeneous density of chirality together with a flow of chirality that is not purely transverse as in the case of a single SP mode. Unlike LSPs, such plasmonic optical lattice fields are {\it delocalized}. This implies that coherent superpositions of SPs constitute original sources of delocalized optical chirality, relevant both in the context of chiral spectroscopy and in the context of optical forces through inhomogeneous chiral densities and uncompensated chirality flows. A central outcome of our analysis lies in the intrinsic plasmonic character of such combined capacities since all pertinent quantities identically vanish in the case of  purely evanescent waves. \\
\indent From an experimental perspective, the delocalized character of the plasmonic optical lattice is interesting since it does not require the capacity to bring chiral molecules at specific locations as it is the case for chiroptical schemes based on LPSs. Additionally, the chiral optical forces, just as their optical chiral sources, spatially extend throughout the interfering region of the optical lattice, opening the possibility to act on large numbers of chiral systems, such as chiral colloidal suspensions \cite{yan2012self,mcpeak2014complex}.
%

\section{Optical chirality}

For a general electromagnetic field $\left( \bcalE(\bor,t), \bcalH(\bor,t) \right)$, the chiral properties of light are described by the density of chirality $K$ and the flow of chirality $\boPhi$ \cite{lipkin1964existence,bliokh2011characterizing,cameron2013optical,canaguier2013mechanical}:
\begin{align}
&K(\bor,t) = \frac{\eps}{2} \bcalE \cdot \left( \nabla \times \bcalE \right) + \frac{\mu}{2} \bcalH \cdot \left( \nabla \times \bcalH \right) \label{K_general} \\
&\boPhi (\bor,t) = \frac{1}{2} \bcalE \times \left( \nabla \times \bcalH \right) - \frac{1}{2} \bcalH \times \left( \nabla \times \bcalE \right) \label{phi_general}\end{align}
where $\eps, \mu $ are the (real) permittivity and permeability of the medium with $\eps \mu c^2=1$. These quantities obey a conservation law $\nabla \cdot \boPhi + \partial_t K = 0$, analogous to the conservation equation relating the electromagnetic energy density and flow.  In the particular case of a harmonic field $\left( \bcalE, \bcalH \right) = \Re \left( \bE_0(\bor) e^{-\imath \omega t}, \bH_0(\bor) e^{-\imath \omega t} \right)$, the quantities (\ref{K_general},\ref{phi_general}) become time-invariant, and their expressions can be simplified to \cite{bliokh2011characterizing,canaguier2013mechanical}:
\begin{align}
&K(\bor) = \frac{\omega}{2c^2} \Im \left[ \bE_0 \cdot \bH_0^* \right]  \label{K_harmonic} \\
&\boPhi (\bor) = -\frac{\omega}{4} \left( \eps \Im[\bE_0 \times \bE_0^{*}]  + \mu  \Im[\bH_0 \times \bH_0^{*}] \right)   ~ .\label{phi_harmonic}\end{align} 

It is interesting to note that the degree of chirality  of a generic harmonic field, defined as the ratio between the chirality density and the time-averaged energy density $W(\bor)=\varepsilon \|\bE_0\|^2/4 +\mu \|\bH_0\|^2/4$, turns out to be strictly bounded \cite{bliokh2011characterizing} with 
 \begin{align}\label{K_bound}
 \left| \frac{c K(\bor)/\omega}{W(\bor)}\right| \leq 1
 \end{align}
the equality being reached for circularly polarized light (CPL). Provided harmonicity, the bound (\ref{K_bound}) is general and applies both to near and far fields. This normalization thus appears as an apposite measure of the degree of chirality for a light field. It emphasizes indeed that the dramatic increase in chirality density reported at the level of plasmonic nanostructures is directly connected with large field enhancements occurring because of specific local concentrations of electromagnetic energy on the nanostructures.  \\
 \indent When interacting with a chiral dipole, characterized by a non-zero mixed electric-magnetic complex polarizability $\chi $, a chiral electromagnetic field exerts a time-averaged chiral force, which can be separated into reactive and dissipative chiral components \cite{canaguier2013mechanical}:
 \begin{align}\label{chiral_force}
 \boF_{\chi} = \Re[c\chi] \frac{c}{\omega}  \nabla K + \Im[c\chi] \frac{2}{c}  \left( \boPhi - \frac{1}{2} \nabla \times \PI \right) ~ ,
\end{align}
with the Poynting vector $\PI(\bor)=\langle {\bf S}(\bor,t) \rangle ={\rm Re}[\bE_0\times\bH_0^{*}]/2$. The total optical force exerted on the chiral dipole is therefore the sum of the chiral force (\ref{chiral_force}) and the usual achiral components that are the gradient force and the dissipative force \cite{canaguier2013mechanical}. Eq.(\ref{chiral_force}) shows that the description of light chirality cannot be reduced to  the sole polarization ellipticity:  for instance, a propagating CPL leads to a vanishing reactive chiral force because of its homogeneous chiral density. \\
\indent From this perspective, it is interesting to stress that near-field inhomogeneities are expected to lead to strong gradients of chirality density and therefore to important reactive chiral forces. Contrasting with the problem of the absorption rate of a chiral molecule that simply involves the density itself, a discussion  on optical forces gives an additional argument in favor of the near field where field localization can induce large inhomogeneities in the chirality density.  As a clear potential, we note that the chiral force in Eq.~(\ref{chiral_force}) changes sign when switching between two enantiomeric forms of the chiral object. This unique property is at the core of the definition of new optical schemes for enantioseparation \cite{canaguier2013mechanical,cameron2014discriminatory,cameron2014diffraction,tkachenko2014optofluidic}. 

\section{The generic surface plasmon}

We first consider an SP mode propagating at a metal-dielectric interface located at $z=0$.  The dielectric medium lies in the ($z>0$)-half space, with $\eps = n^2 \eps_0, \mu=\mu_0$, where $n \geq 1$ is the real-valued refractive index. The metal medium lies in the ($z<0$)-half space and is defined by a complex permittivity $\varepsilon_m$ that is frequency-dependent. The SP mode propagates in the $x$-direction and is characterized by a complex wave vector $(k,0,q)^t$ that obeys a specific dispersion relation with $k=\omega / c \sqrt{\varepsilon_m / (\varepsilon_m + \varepsilon)}$ and $k^2+q^2=(\omega /c)^2$.

As far as chirality is concerned, the disappointment regarding such a generic SP mode directly stems from its transverse magnetic (TM) nature which implies, as already recognized, an identically vanishing chiral density \cite{bliokh2012transverse}. Indeed, the magnetic field $\bH_0 = H_0~ e^{\imath kx}e^{\imath qz} \left(0, 1, 0 \right)^t$ gives an elliptical SP electric field $\bE_0 = E_0~ e^{\imath kx}e^{\imath qz} \left( \tq, 0, -\tk\right)^t$ which is normal to $\bH_0$, and where $\tk=kc/\omega$ and $\tq=qc/\omega$ are respectively the dimension-less longitudinal and transverse wave vector components and $H_0=E_0\sqrt{n^2\eps_0/\mu_0} $. The plasmonic electric ellipticity leads to a non-zero flow of chirality which can be easily obtained from Eq.~(\ref{phi_harmonic}) as $\boPhi=I_0 \Phi(x,z)\hat{\bf y}$ with
\begin{align}\label{phi_plasmon}
\Phi(x,z) = \frac{\omega}{2c}  e^{-2k''x-2q''z} ~\Im[\tk \tq^{*}]  ~,
\end{align}
and where $I_0=E_0 H_0^*$ is the intensity of the field.

The flow of chirality is transverse to the direction of propagation, a property that directly comes from the evanescent character of the plasmonic field. As recently shown, this is responsible for transverse torques exerted on dissipative nanospheres \cite{bliokh2014extraordinary,canaguier2013transverse}. But when considering chiral forces, the flow of chirality turns out be exactly compensated by the curl of the time-averaged Poynting vector of the generic SP mode with $\boPhi = \nabla \times \PI / 2$. This leads to a vanishing dissipative part of the chiral force given in Eq.~(\ref{chiral_force}) and therefore to the important conclusion of the absence of chiral forces induced on chiral dipole immersed in the near field of a propagating generic SP mode. This conclusion actually holds for any transverse magnetic wave, either propagating or evanescent.

\section{The plasmonic optical lattice}
We have however realized that the situation of the coherent superposition of two SP modes can totally changes the landscape and opens new opportunities. First, as we now demonstrate, the near-field optical lattice formed by the intersection of two SP modes does carry an inhomogenenous chiral density together with a chirality flow this time uncompensated by the curl of the Poynting vector resulting from the plasmonic interference. 

\begin{figure}[htbp]
\centering{
\includegraphics[width=0.5\textwidth]{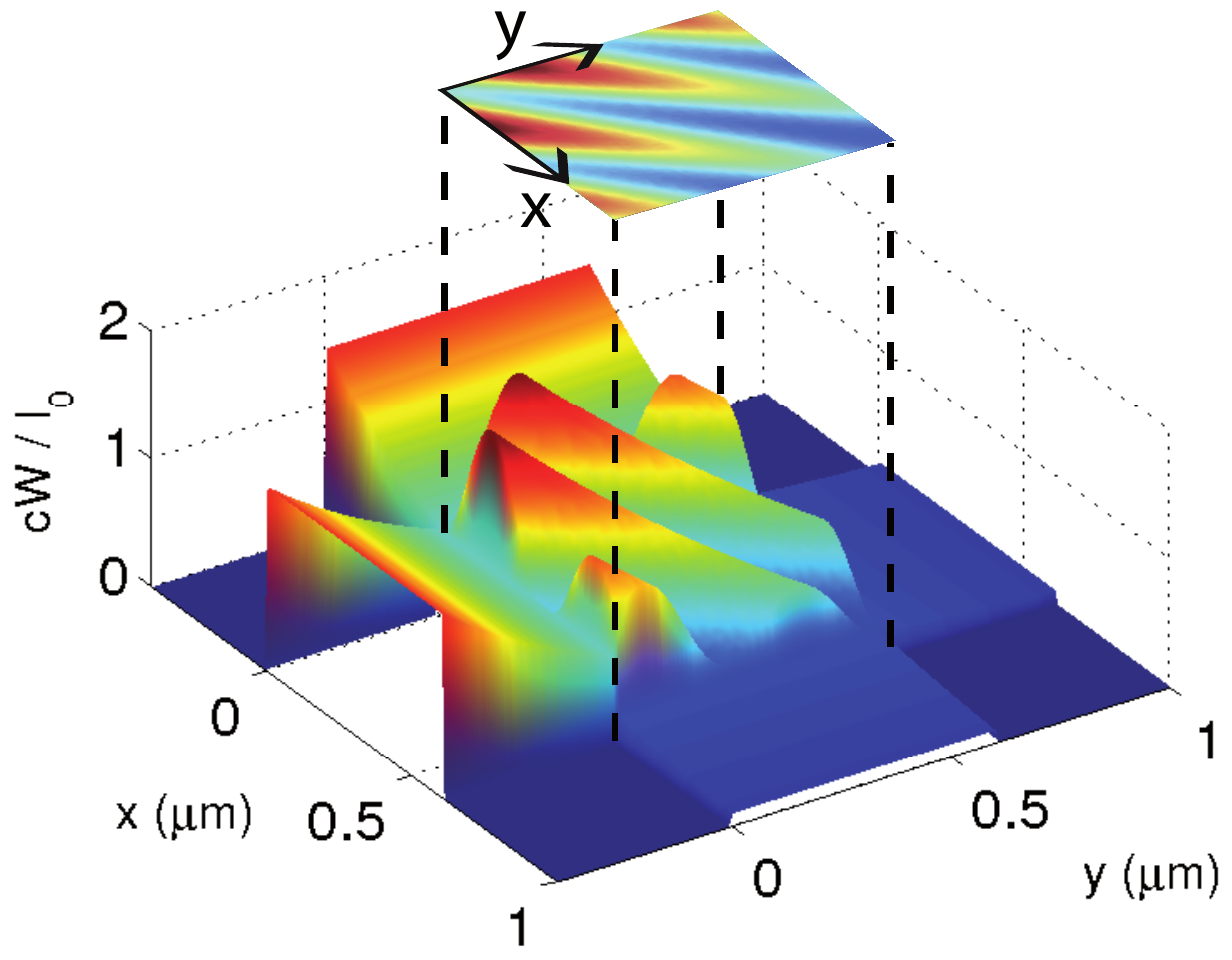}}
\caption{Sketch of the plasmonic optical lattice:  two SPs are propagating along the two orthogonal $x$ and $y$ directions, on the same gold-water interface. The figure displays the total electromagnetic energy $W(x,y)$ on top of the metallic layer ($z=0^+$), normalized by $I_0/c$, with a vignetting of the two surface plasmons to show their respective propagation directions and reveal the interfering region, shown in the upper inset. In this evaluation, a $540$ nm wavelength has been chosen in order to increase the plasmonic longitudinal exponential decay making the two orthogonal propagation directions more visible.}
 \label{fig_K}
\end{figure}

We consider the coherent superposition of two orthogonal SPs propagating on the same metal-dielectric interface with the same angular frequency $\omega$, as sketched in Fig. \ref{fig_K}:  the two SPs are propagating along the $x$- and $y$- directions and are intersecting perpendicularly to each other. The resulting near field in the upper dielectric medium then writes as the superposition of the two generic SPs:
\begin{align}\label{2DSP_field}
&\bE_0 = ~ e^{\imath qz} \left(  \tq E_1 e^{\imath kx} , -\tq E_2 e^{\imath ky} , \tk \left( E_2 e^{\imath ky} - E_1 e^{\imath kx} \right) \right)^t \nonumber \\
&\bH_0 = e^{\imath qz} \left( H_2 e^{\imath ky} , H_1 e^{\imath kx} , 0  \right)^t ~ ,
\end{align}
where the complex amplitudes of the two SPs are related through $H_1 = \sqrt{n^2\eps_0/\mu_0} E_1$ and $H_2 = \sqrt{n^2 \eps_0 / \mu_0} E_2 $. For simplicity, we label the intensities of the uncoupled SP fields as $E_1 H_1^{*}=I_1, E_2 H_2^{*}=I_2$ and of the coupled fields as $E_1 H_2^{*} = H_1 E_2^{*} = I_{1,2} e^{\imath \theta}$, $\theta$ being the fixed phase difference between the two surface plasmons at the origin.

The presence, in the interference process, of a coupling term between the fields of the two intersecting SPs leads to the first important result that the chirality density $K(\bor)$ associated with the plasmonic optical lattice is non-zero and inhomogeneous. It can be evaluated directly from Eq.~(\ref{K_harmonic}) as:
\begin{align}\label{2DSP_K}
K(\bor) = \frac{\omega I_{1,2}}{c^2} e^{-k'' (x+y)-2q'' z} \tq' \sin \phi ~ ,
\end{align} 
where $\phi(x,y)=k'(x-y) + \theta$ is the local phase difference between the two waves.

\begin{figure}[htb]
\centering{
\includegraphics[width=0.4\textwidth]{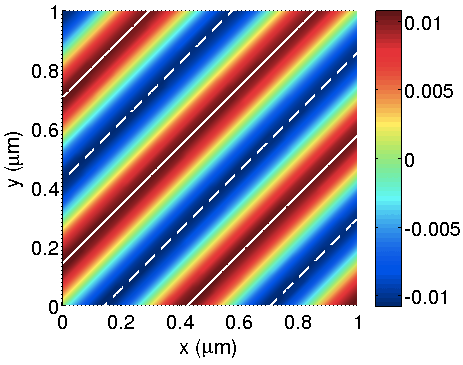} 
\includegraphics[width=0.32\textwidth]{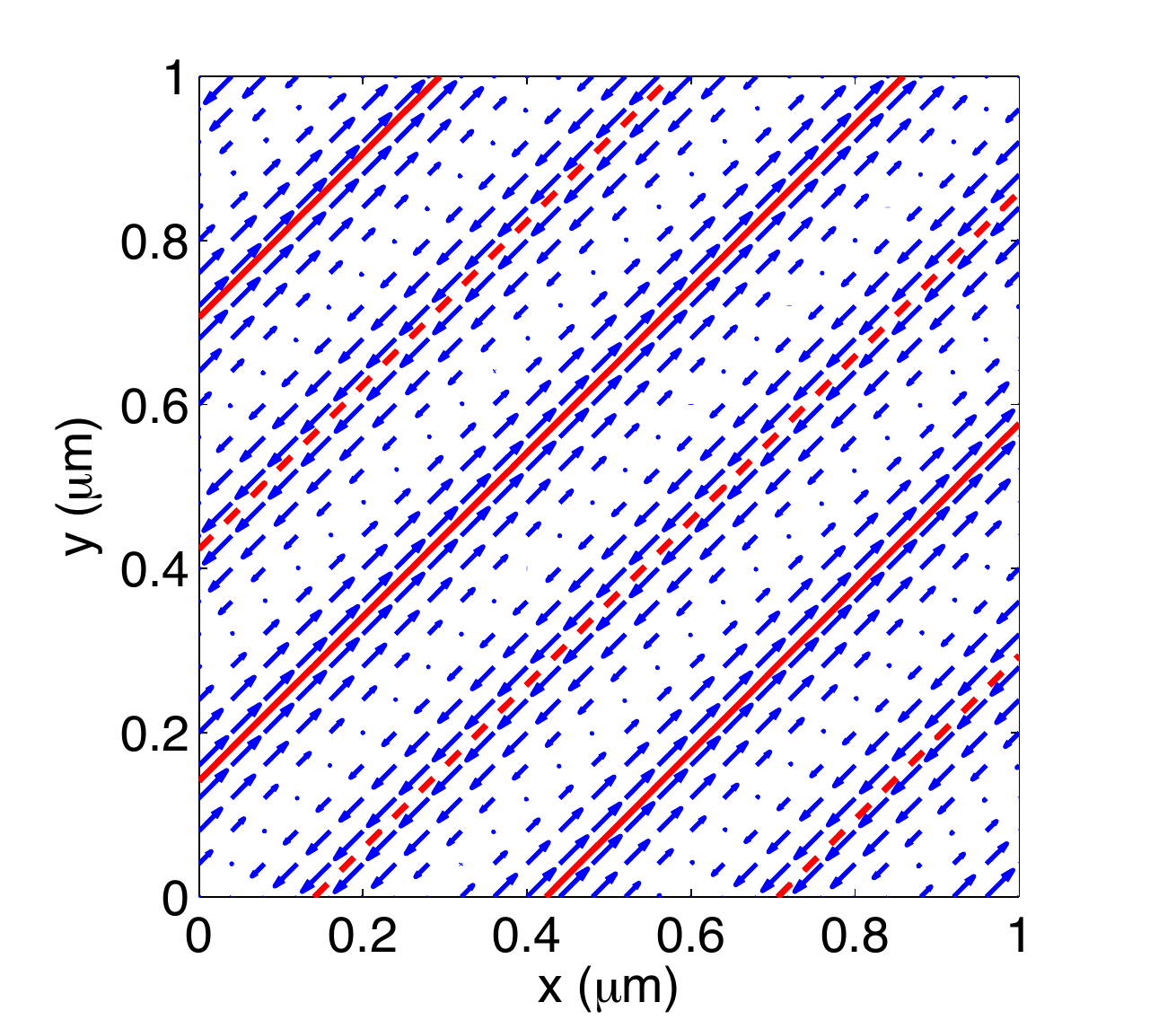}~ \hspace{1.2cm} ~}
\caption{(a) Density of chirality $K(\bor)$ as a function of position $(x,y)$ on top of the metallic layer ($z=0$),  normalized by $\omega I_{1,2}/c^2$. The chirality density is  minimized along lines $L_p^{-}$ (dashed white lines) and maximized along lines  $L_p^{+}$ (solid white lines). The reactive chiral force defined from the gradient of $K(\bor)$ will essentially be directed perpendicularly to the $L_p^{\pm}$ lines. (b) Direction and relative amplitude of the $x$ and $y$ components of the vector field $\boPhi-\nabla\times \Pi /2$ given in Eq. (\ref{2DSP_phiO}), in the $z=0$ plane. The lines $L_p^{\pm}$ are displayed in red for comparison with panel (a). Both panels correspond to the scheme presented in Fig.~\ref{fig_K} where two surface plasmons of identical intensities are launched perpendicularly from each other at a $780$ nm wavelength on a gold-water interface, with no phase difference at the origin, i.e. with $\theta=0$.}
 \label{fig_xy}
\end{figure}

As a consequence, a reactive chiral force is induced (first term in Eq. (\ref{chiral_force})) at the level of which the normalized chiral density $K(\bor) c/\omega$ can be interpreted as a genuine potential energy density  which pushes the chiral dipole towards areas where $K$ is maximized or minimized, depending on the sign of $\Re[c \chi]$. On the $z=0$ plane for instance, the $\sin \phi = \sin[k'(x-y)+\theta]$ term has identical values along lines $y=x-a$ with $a$ a real constant, as illustrated in Fig.~\ref{fig_xy} (a). 
Keeping in mind that $\tq'<0$ for a metal-dielectric interface, $K$ will be minimized for lines $L_p^{-}$ along which $\sin\phi=1$ ($k' a+\theta = \frac{\pi}{2} + 2p\pi$, $p$ an integer) and  maximized for lines $L_p^{+}$ along which $\sin\phi=-1$ ($k'a+\theta = -\frac{\pi}{2} + 2p\pi$). This is presented in Fig.~\ref{fig_xy} (a) with dashed and solid lines, respectively. The density modulation is superimposed to the long range amplitude decrease for increasing values of $(x+y)$ due to the plasmonic damping factor $e^{-k'' (x+y)}$ in Eq. (\ref{2DSP_K}), not discernible in Fig.~\ref{fig_xy} because the chosen wavelength $\lambda=780 nm$ yields a too small damping.
 
The second central outcome is related to the chirality flow $\boPhi$ which can be obtained directly from Eq.~(\ref{phi_harmonic}) and reads as a superposition $\boPhi=\boPhi_1+\boPhi_2+\boPhi_{12}$ of the chirality flow associated with each SPs and of a coupling between the two SP modes. From Eq.(\ref{phi_plasmon}), we have $\boPhi_1=I_1 \Phi(x,z)\hat{\bf y}$ and $\boPhi_2=-I_2 \Phi(y,z)\hat{\bf x}$. The coupling term
\begin{align}\label{2DSP_flux}
\boPhi_{12}=\frac{\omega  I_{1,2}}{2 c} e^{-k''(x+y)-2q'' z} \left( \begin{array}{c} \Im \left[ \tk \tq^{*} e^{\imath \phi}\right] \\ \Im \left[ \tk^{*} \tq e^{\imath \phi}\right] \\ \left( 1 + |\tq|^2 \right) \sin \phi \end{array} \right)
\end{align}
surprisingly has a non-zero component in the $z$-direction which first factor comes from a magnetic ellipticity induced by the coherent superposition. Importantly too, the direction of this component of the chirality flow oscillates as a function of  the local phase difference $\phi$ between the two waves. At this stage, these properties remain both for propagative waves, taking real $k$ and $q$ wave vectors, and for a purely evanescent fields with real $k$ but pure imaginary transverse wave vector $q$. \\
\indent In the force perspective, it is clear that the first two terms $\boPhi_1$ and $\boPhi_2$ are identically compensated by the curl of the associated time-averaged Poynting vectors $\PI_1$ and $\PI_2$. But $\boPhi_{12}$ is only altered, and not canceled, by the same curl related to the SP coupling. We are thus left with
\begin{align}\label{2DSP_phiO}
\boPhi - \frac{ \nabla \times \PI }{2}= \tq'\frac{\omega  I_{1,2}}{2 c} e^{-k''(x+y)-2q'' z} \left( \begin{array}{c}   \Im \left[ \tk  e^{\imath \phi}\right] \\   \Im \left[ \tk^{*} e^{\imath \phi}\right] \\ 2  \tq'  \sin \phi \end{array} \right) ~,
\end{align} 
which implies that the plasmonic optical lattice does yield a non-zero dissipative chiral force.  The longitudinal $x$ and $y$ components of this vector field are presented in Fig. \ref{fig_xy} (b) as a function of position in the $z=0$ plane. It reveals a very different in-plane dynamics as compared with the reactive chiral force since the dissipative chiral effects are alternatively oriented along the $L_p^{\pm}$ lines, and not perpendicular to as in the case of the reactive chiral force. 

\begin{figure}[htbp]
\centering{
\includegraphics[width=0.38\textwidth]{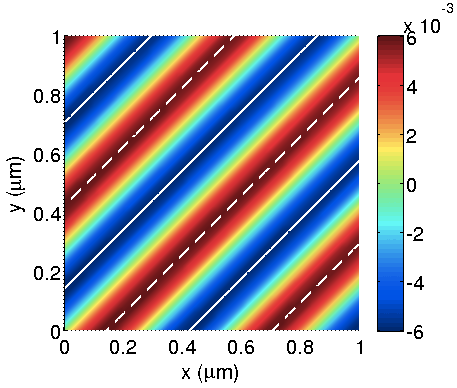}
\includegraphics[width=0.38\textwidth]{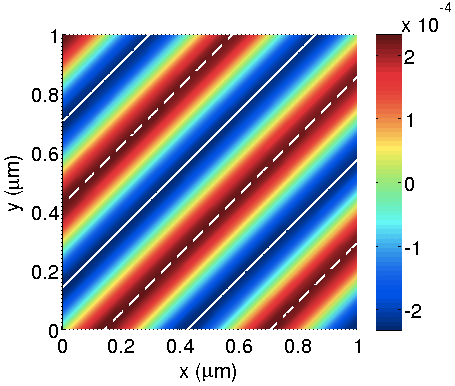}}
\caption{(a) Gradient of the chirality density in the $z$-direction, normalized by $\omega^2 I_{1,2}/c^3$.  (b) $z$-component of the vector field $\boPhi-\nabla\times \Pi /2$ in the $z=0$ plane, normalized by $\omega I_{1,2}/2c$. These patterns correspond to the scheme of orthogonal surface plasmons involved in Fig.~\ref{fig_xy}. The lines $L_p^{\pm}$ are presented for comparison with Fig.~\ref{fig_xy}. Panels (a) and (b) can be quantitatively compared since both figures are equivalent to a force density normalized by $\omega I_{1,2}/c^2$.}
 \label{fig_z}
\end{figure}

Simultaneously, because of the evanescence of the plasmonic field in the $z$-direction, the chiral forces have strong components in the $z$-direction which will lead to vertical chiral forces. Such force components are displayed in Fig. \ref{fig_z} (a) and (b) respectively for the reactive and dissipative components. 
Because the $z$-derivative of Eq. (\ref{2DSP_K}) leads to a $(-2q'')$ factor, the $z$-component of the reactive chiral optical force turns out to be opposed to the density of chirality, as easily seen by comparing the two panels (a) of Figs \ref{fig_xy} and \ref{fig_z}. The $z$-component of the dissipative chiral force is shown in Fig. \ref{fig_z} (b) and reveals an identical pattern than the one obtained from the chirality density gradient. Due to the strong amplitude gradient in the $z$-direction, the reactive part of the force density presented in in Fig. \ref{fig_z} (a) is 20 times larger than the dissipative part presented in in Fig. \ref{fig_z} (b) for the considered wavelength. Nevertheless, when considering the total chiral force one has to keep in mind that these force densities are modulated by the real and imaginary parts of $(c\chi)$ which generally depends on the optical frequency. 

Finally, we stress that both quantities presented in Eqs.~(\ref{2DSP_K},\ref{2DSP_phiO}) that characterize the two components of the chiral force are proportional to the real part of the wave vector $q$. As an important consequence, an evanescent optical lattice, made from the intersection of two purely evanescent fields will not lead to any chiral optical forces. This underlines a fundamental difference between plasmonic fields and evanescent fields such as implemented in TIRF-based studies. In particular, this clearly points to the fact that plasmonic optical lattices are specifically promising when aiming at deracemization schemes exploiting chiral near field forces. The reactive chiral force landscapes displayed in Figs \ref{fig_xy} (a) and \ref{fig_z} (a) turn out particularly appealing for optical chiral separation since each enantiomer will be pulled down towards the interface ($F_z<0$) within the attractive regions shown in Fig. \ref{fig_xy} (a) and pushed away ($F_z>0$) over repulsive regions. 
%

\section{Conclusions}
We have shown that a near field with a rich optically chiral near field can be realized by forming a plasmonic optical lattice with two normally intersecting surface plasmons. In this configuration, the coupling between the two coherent surface waves is the source for inhomogeneous optical chirality and therefore can induce chiral optical forces on chiral objects. Because the strong amplitude and phase gradients associated with plasmonic fields have proven to lead to enhanced momentum transfers in the near field, the generation of near field chirality from plasmonic optical lattices opens promising perspective in the context of chiral manipulation and separation schemes. This is all the more true given that the amplitude and sign of both real and imaginary parts of $\chi$ depend on the frequency of the near field thus leading to a great variety of chiral dynamical effects that can be generated at the level of such plasmonic optical lattices. Our results therefore add to the chiroptical toolkit a new class of field excitations in the form of plasmonic optical lattices. 

\bibliography{biblio_chiralplasmon}

\end{document}